\documentstyle[12pt]{article}
\newfont {\xx} {cmti10}

\def\be{\begin{equation}}

\def\ee{\end{equation}}
\def\bea{\begin{eqnarray}}
\def\eea{\end{eqnarray}}

\begin{document}
\begin{titlepage}
\begin{flushright}
{OUTP-97-34-P}\\
{May 1998}
\end{flushright}
\vskip 0.5 cm

\begin{center}
            {\large{\bf Aspects of String unification}}
            \vskip 0.6 cm
{{ Witold Pokorski 
\footnote{E-mail address: W.Pokorski1@physics.ox.ac.uk}, Graham G. Ross
\footnote{E-mail address: Ross@thphys.ox.ac.uk}
}}\\
\vskip 0.3 cm
{Department of Physics, Theoretical Physics,\\
University of Oxford, 1 Keble Road, Oxford OX1 3NP}
\date{}
\end{center}
\vskip 2 cm
%\maketitle
\abstract{We consider the phenomenological implications of a class of
compactified string theories which naturally reproduces the flavour
multiplet structure of the Standard Model. The implications for gauge
unification depends on which of three possibilities is realised for
obtaining light Higgs multiplets. The more conventional one leads to
predictions for the gauge couplings close to that of the MSSM but with an
increased value of the unification scale. The other two cases offer a
mechanism for bringing the prediction for the strong coupling into agreement
with the measured value while still increasing the unification scale. The
various possibilities lead to different expectations for the structure of
the quark masses.}
\end{titlepage}
\setcounter{footnote}{0}
\section{Introduction}

Superstring theory offers us the exciting prospect of a complete unification
of all the fundamental interactions including gravity. A preliminary study
of its general implications for low-energy structure is quite encouraging.
The heterotic string has a gauge symmetry which contains the Standard Model
gauge group. In a definite string vacuum the number of massless modes (on
the scale of the Planck mass) is determined so there is the prospect of
understanding the origin of three families. There are viable predictions for
the fundamental couplings, such as gauge couplings, even in the absence of a
stage of Grand Unification. However the analysis of superstring
phenomenology has been hampered by the profusion of candidate 4-dimensional
string theories, corresponding to different string vacua. In this paper we
attempt to make some progress by identifying a class of string models that
most closely generates the Standard Model structure. While this is certainly
not exhaustive of the possible realistic string theories it does, we think,
represent a most promising class. Interestingly, within this class one may
draw some quite general conclusions about the nature of the low-energy
theory and the predictions for the parameters of the Standard Model.

To set the stage for this discussion let us briefly review the most
significant features of the Standard Model. A significant feature is that
the observed quark and lepton states of the Standard Model transform as
singlets or as the fundamental representation of the Standard Model gauge
group. Even more noticeable is that the quarks and leptons fill out complete
representations of $SU(5)$ (and $SO(10)$?) even though the low-energy gauge
group is not $SU(5)$. Moreover the $SU(5)$ assignment of Standard Model
states offers an immediate explanation for the left-handed nature of the
weak interactions for both the quarks and leptons. However this symmetric
view is somewhat spoilt by the sector responsible for spontaneous symmetry
breaking. The Higgs doublet is the only representation of matter fields in
the Standard Model spectrum that does not have its $SU(5)$ partners
(additional $SU(2)$ singlet d quarks).

Let us first consider how these features are accommodated in Grand Unified
theories. Clearly the $SU(5)$ assignments strongly suggest an underlying GUT
containing $SU(5)$. However a GUT does not by itself explain why only the
fundamental representations of the gauge group occur. Moreover in Grand
Unified theories the partial GUT representation of the Higgs must be
explained and this gives rise to the doublet-triplet splitting problem,
namely the need to give the partners of the Higgs a mass while leaving the
Higgs doublets light. Solutions to this problem exist but typically require
a large number of additional Higgs fields and a somewhat unconvincing and
complicated set of interactions. In addition Grand Unified theories (GUTs)
must be protected against the radiative coupling of the large Grand Unified
scale to the low electroweak breaking scale - the hierarchy problem. Its
solution requires a low energy stage of supersymmetry. This leads to the
MSSM, the supersymmetric extension of the Standard Model. Grand Unification
of couplings plus the radiative corrections of the MSSM lead to the
remarkably successful prediction for the ratio of gauge couplings at low
energy provided the $U(1)$ normalization is chosen to be that given by $%
SU(5) $.

How is this unification picture changed in superstring theories \cite{gsw}?
Certainly the string theory has an enlarged symmetry; in the heterotic
string $E_{8}\otimes E_{6}$ $\cite{heterotic}$. Although this symmetry may
be broken at the compactification scale a residue of it may persist in the
light spectrum even if there is no Grand Unified group. This may explain the
pattern of quark and lepton supermultiplets. Moreover, if the string theory
is built from level-1 Ka\v{c}-Moody level theories, the representation
content of the theory is restricted offering an explanation to the question
why only low lying representations of quark and leptons are observed. As we
shall discuss in more detail shortly the string offers an elegant
explanation for the doublet triplet splitting problem in the case there is 
{\it no} stage of Grand Unification below the Planck scale. In this case
there is no symmetry reason demanding there should be partners to the Higgs
thus finessing the doublet-triplet problem. However the string now has to
explain why the quarks and leptons still come as complete $SU(5)$
representations; we shall discuss how this happens naturally in a class of
string theories.

Unification of couplings is a prediction of string theories even without a
Grand Unified group below the Planck scale. In this case it is a residue of
the underlying gauge symmetry before compactication. Taken with a stage of
low-energy supersymmetry needed to solve the hierarchy problem this may
provide an alternative explanation for the success of gauge coupling
unification. Further the string makes an important additional prediction
which goes beyond Grand Unification, namely it determines the unification
scale in terms of the Planck scale. If this could be checked it would
provide the first quantitative test of the unification of the strong,
electromagnetic and weak forces with gravity which the superstring provides.
Preliminary indications are promising in that the scale of unification
predicted by continuing the gauge couplings up in energy from the laboratory
scale assuming the minimal supersymmetric Standard Model (MSSM) spectrum
gives a very large scale of unification of $O((1-3)10^{16}){\rm GeV}$)) not
so far from the Planck scale which sets the scale for the string prediction.
In the case of the (weakly coupled) heterotic string the detailed prediction
is \cite{kap,dienes} 
\begin{equation}
M_{string}\approx g_{string}\times (5.2\times 10^{17}{\rm GeV})\approx
3.6\times 10^{17}{\rm GeV}  \label{su}
\end{equation}
Although this is relatively close to the gauge unification scale it is
still a factor of 10-30 too high. One should remember that the Grand
Unified scale is the argument of the logarithm and thus, in order to
get it correctly, one has to work to very high precision in the
coupling constant. Possible problems in determining this scale are :
\begin{itemize}	
\item SUSY threshold effects -  It seems that although these could make the factor of 20 difference this only can be achieved with a very peculiar supersymmetry spectrum at low energies in which the gluino is lighter than the Wino \cite{A13}.
	
\item Additional heavy states - A second possibility is that the theory at compactification is not just the Standard Model but is still Grand Unified. In this case the string prediction for the unification scale applies to the unification of couplings of 

the Grand Unified theory. While possible it limits the predictive power of the string. It may also be that there are additional heavy states which just conspire to change the unification scale. Examples of this are given in \cite {DF}.
\item String threshold effects - Another possibility that has been
explored are string threshold effects which in a given string theory
are calculable \cite{A15} and amount to including the effect of the heavy Kaluza Klein modes which are themselves split when the Grand Unified theory is broken. While it is possible to construct examples in which these effects are large \cite{ilr}, more t
ypically they are too small to explain the discrepancy  \cite{dienes}. However it has recently been shown that threshold effects associated with Wilson line breaking can, for reasonable values of the moduli, generate effects easily large enough to explain

 the discrepancy \cite{NS}. While such effects typically change the relative evolution of the couplings and hence spoil the success of gauge unification, it has been shown how, for a subset of models, only the scale of unification may be changed as desire
d. Of course one is left with the question why this subset of models is selected.

\item Large gauge coupling - A mechanism which automatically leads to just a change in the scale of  unification was proposed by Witten \cite{witten}. He observes that, in the absence of large corrections of the type just discussed, the string prediction 

which is based on the assumption of a weakly coupled heterotic string breaks down. This is because the fit to the values of the low energy gauge couplings and Newton's constant require the ten dimensional string coupling to be very {\it large}. Thus Witte
n argues one should do the string calculation for the  case the 10
dimensional gauge coupling is large in the heterotic string theory (This is the M
theory limit \cite{horwit}). In this case he found the prediction for the gauge unification scale changes and is dependent on a new parameter, the compactification scale associated with the eleventh dimension of M-theory.
\end{itemize}

In this paper we wish to select from candidate string compactifications
those that can explain the features of the effective low-energy theory and the unification predictions just discussed. Given the plethora of
candidate string theories this seems a very tall order but we will argue
that the most predictive string theories must belong to a very small
subclass. Let us first consider the prospects for making a superstring
prediction for the parameters {\it including} the gauge unification scale of
the effective low-energy theory descending from the string. The class of
string theory which gives such predictions for the unification scale of the
Standard Model gauge couplings is quite restricted. In it there {\it must not%
} be a stage of Grand Unification {\it below} the compactification scale
because the prediction refers to the unification scale for the couplings of
the gauge group at the compactification scale and these must just be those
of the Standard Model. Even without such a stage of Grand Unification the
gauge couplings are related by the underlying string symmetries \cite
{ginsparg}. This is readily achieved with the required  ($SU(5)$-like)
normalization for the $U(1)$ gauge factor in string theories in which the Ka%
\v{c}-Moody level used in the string construction is restricted to be level-1%
\footnote{%
See \cite{dienes} for a general discusssion of the other possibilities.}. An
advantage of using such level-1 string theories is that they have a
restricted multiplet structure with only fundamental representations for the
chiral supermultiplets. Encouragingly this is in agreement with the representation
content of the Standard Model in which the quarks and leptons transform as
singlets or as the fundamental representation of $SU(3)\otimes SU(2)$. Given
this we concentrate on such level-1 theories. (Note that level-1 theories 
{\it cannot} have a stage of Grand Unification with a simple gauge group
{\it below} the compactification scale because it is necessary to have larger
Higgs representations to break such simple Grand Unified groups to the
Standard Model.$)$

We further restrict our string compactifications to those that offer an
explanation of the $SU(5)$-like multiplets of quark and lepton fields even
though the gauge group is not-Grand Unified below the compactification
scale. To do so we need some relic of an underlying string gauge symmetry to
persist in the low energy spectrum. Thus we  concentrate on exploring in
some generality the low-energy structure of a class of such level-1 string
theories which correspond to a compactified heterotic string theory \cite
{candelas}\footnote{%
This does not cover all 4-dimensional string theories for some cannot be
viewed as a compactification from 10 dimensions. However in them there is no
obvious underlying extended gauge symmetry which requires the $SU(5)$
patterns observed.}. To make a start in this we note that the underlying
symmetry of the 10 dimensional heterotic string includes an $E_{8}\otimes
E_{6}$ or $SO(32)$ symmetry\footnote{%
Henceforth we will only consider the phenomenologically interesting $%
E_{8}\otimes E_{6}$ case where $E_{6}$ contains the Standard Model group.}.
This symmetry has to be broken in the 4 dimensional theory but a residue of
the underlying Grand Unification will remain offering an explanation for the 
$SU(5)$ content of the matter fields of the Standard model. As we
mentioned before, the mechanism
for breaking the 10 dimensional gauge symmetry cannot be the usual Higgs
mechanism because the Higgs content is restricted in a level-1 string theory
and is insufficient to break the symmetry down to the Standard Model gauge
group. However it has been shown that the necessary breaking can occur
through Wilson line breaking \cite{wilson} in which flux is trapped in
non-simply-connected manifolds. Such non-simply-connected manifolds can
arise through the modding out of the manifold by a freely acting discrete
group with the Wilson lines forming a representation of the group. Indeed
such Wilson line breaking can break the gauge symmetry down to just that of
the Standard Model. We will consider the class of models in which this is
the case as we wish to explore the possibility that the string prediction
for the gauge coupling unification scale applies directly to the Standard
Model couplings.

The implications of Wilson line breaking for the chiral multiplet content is
particularly interesting because such breaking does not reduce the chiral
content of the theory. If in the $E_{6}$ visible sector there are $(n+m)$
left-handed chiral superfields transforming as the $27$ representation ( $%
\left( 2,1\right) $ harmonic forms) and just $m$ left-handed chiral
superfields transforming as the $\overline{27}$ representation ($(1,1)$
harmonic forms) there is an excess of $n$ ``chiral '' superfields with the
necessary states to include the (massless) generations of the MSSM. Wilson
line breaking does not affect the chiral structure of the theory and so in
the compactified theory we are guaranteed to have $n$ left-handed chiral
superfields with the same gauge content as occurs in complete $27$
representations of $E_6$ even though the gauge group is just that of the
Standard Model.

In addition to these supermultiplets there will be further states left light
after Wilson line breaking, related to the original $m(27+\bar{27})$
superfields in the theory without Wilson line breaking. There are two
possibilities depending on whether, before modding out by the discrete
group, they correspond to the $m_{s}(27+\bar{27})$ fields which are singlet
representations under the discrete group or they correspond to the $%
m_{ns}D(27+\bar{27})$ fields which belong to non-singlets representations of
dimension D under the discrete group. In the latter case, after modding out,
there are left light states with the same gauge content as in $m_{ns}$ {\it %
complete} $(27+\bar{27})$ representations whether or not there is Wilson
line breaking. In the absence of Wilson line breaking these are just the
components of the discrete group non-singlet representations which are left
invariant under the discrete group transformation. In the case there are
non-trivial Wilson loops associated with the singularities introduced by the
modding out by the discrete group the light states do not correspond to the
same original multiplets in the theory before Wilson line breaking. The
reason for this is that the light fields must be singlets under the combined
transformation of the Wilson line group element and the discrete group
element. Thus the components of the $m_{ns}D(27+\bar{27})$ which are left
invariant by the Wilson line group element must also be left invariant by
the discrete group and thus correspond to the components of the
corresponding multiplets in the compactification without Wilson line
breaking. However the components of the $m_{ns}D(27+\bar{27})$ which
transform non-trivially under the Wilson line group element must also
transform under the discrete group in such a way that they are invariant
under the combined transformation. Hence they correspond to {\it different}
multiplets from the case without Wilson line breaking. Rather they
correspond to the appropriate fields with non-trivial discrete
transformations in the original manifold before modding out. The same is
true of the components of the $n$ $27$s family of fields discussed above.

To summarize, it is necessary to break the level-1 heterotic string theory
by Wilson lines. Despite the fact this may break the symmetry down to a
non-Grand Unified group, perhaps just the Standard Model group, the $n$
generations have the same gauge content as in complete 27s. The $SU(5)$
content of a $27$ is $10+\overline{5}+1+5+\overline{5}+1$. The ($\nu _{R}$)
singlet and the components of the latter ``vectorlike'' pair of $5+\overline{%
5}$ can acquire $SU(3)\otimes SU(2)\otimes U(1)$ invariant masses so they
are likely to be very heavy. Thus we are left with the conclusion that the
light families fill out complete $SU(5)$ representations just as is
observed. Indeed with the right-handed neutrino component, $\nu _{R}$,
included these families fill a $16$-plet of $SO(10)$. All of this is a relic
of the underlying $E_{6}$ symmetry of the heterotic string which persists in
the spectrum below the compactification scale. In addition we expect new
light states below the compactification scale with gauge representation
content equivalent to $m_{ns}$ complete $27+\bar{27}$ representations.
However these too can acquire $SU(3)\otimes SU(2)\otimes U(1)$ invariant
masses so they are likely to be very heavy.

So far we have discussed light fields after compactification which make up
complete $E_{6}$ multiplets. Their appearance is encouraging as they neatly
explain the existence of GUT multiplets in a compactified theory without a
GUT gauge group. However there remains the troublesome question of
explaining the existence of the Higgs supermultiplets which do not belong to
a light GUT multiplet. With Wilson line breaking there is an excellent
mechanism capable of accommodating the Higgs fields. To see this consider
now the fate of the $m_{s}(27+\bar{27})$ fields which are singlet
representations of the discrete group. For them only those components which
are singlets under the Wilson line group element are left light under Wilson
line breaking. Thus there is an immediate reason for ``split'' multiplets to
arise. Moreover, as the Wilson lines must provide a representation of the
discrete group there is only a finite number of possibilities for the Wilson
line group elements so the possible spectra are limited and the probability
it will lead to just the doublet components left light is non negligible.
Indeed, as we discuss below, a complete classification of Wilson line
breaking shows that there are choices which do just leave the Higgs double
components of the ($27+\bar{27}$) light.

Thus we see, in the case of Wilson line breaking, the light multiplet
structure after compactification consists of light left-handed chiral
superfields which have the same multiplet content as $(n+m_{ns})$ complete $%
27$s and $m_{ns}$ complete $\bar{27}$s of $E_{6}$ even though the gauge
group may be much smaller than $E_{6}$ and need not be a Grand Unified group
at all. In addition there are $m_{s}$ split multiplets consisting of those
components of the $(27+\bar{27})$ which are singlets under the Wilson line
group elements.

The rest of this paper is devoted developing the implications for low energy
physics of this class of superstring theories. In Section 2 we give a more
detailed discussion of the implications of Wilson line breaking. In Section
3 we apply Wilson line breaking to the doublet triplet splitting problem and
show how it leads to two distinct possibilities. Section 4 considers the
phenomenological implications for the class of string model under
consideration, both for the unification of gauge couplings and the
unification scale and for the third generation masses. Section 5 presents
our conclusions.

\section{The massless spectrum after Wilson line breaking.}

Wilson lines may arise in theories compactified on non-simply-connected
manifolds. The usual way of constructing non-simply-connected spaces is to
mod out an initial simply connected manifold, $M$, by some discrete symmetry
group, D. In the absence of Wilson lines, on $M/D$, fields must satisfy 
\begin{equation}
\psi (x)=\psi (fx),  \label{wil}
\end{equation}
where $fx$ is the image of $x$ under the action of D. This of course means
that only the discrete group singlet fields remain after such modding out.
Wilson lines, $U_{f},$ correspond to flux lines trapped in non-contractible
loops $\gamma $ in the non-simply-connected manifold \cite{wilson} 
\begin{equation}
U_{f}=P\exp {\left( -i\int_{\gamma }T^{a}A_{m}^{a}dx^{m}\right) }
\end{equation}
The Wilson line depends on the vacuum configuration of the gauge fields $%
A_{m}^{a}$ in the compactified six dimensions (for $m=4...9$). Thus $U_{f}$
will be a group element of the underlying gauge group $G$ and the set of $%
U_{f}$s give a representation of $D$ in $G$. In their presence the
implication for the light spectrum is much richer. When we go around the
singularity associated with $\gamma $ the wave function acquires a phase $%
U_{f}$ due to the non trivial configuration of the gauge field. In such a
case the condition (\ref{wil}) becomes 
\begin{equation}
U_{f}\psi (x)=\psi (fx).  \label{bc}
\end{equation}
The implications for the light spectrum follow immediately. All of the
components of a chiral supermultiplet on $M$ transforms in a given way under
the discrete group. On the other hand the Wilson line gives {\it different}
phases to the different components (different representations in the
decomposition under the unbroken group) of the initial supermultiplet. As
result, if the state on $M$ is a discrete group singlet, only the component
which does not acquire a phase from the Wilson line remains massless. If the
state on $M$ transforms non-trivially under D picking up an overall phase
one immediately sees from eq($\ref{bc})$ there will be left light only that
part of it which acquires the same phase from the Wilson line.

There is a very important implication of this. States on $M$ which transform
as a representation $R$ under $G$ and which make up a non-singlet
representation of $D$ will give rise to light states on $M/D$ which fill out
a complete representation $R$ of $G$ even though the gauge group on $M/D$ is
the smaller group $G/D.$ Thus if on $M$ the discrete non-singlet states
transform as a $27$s under $E_{6}$ there will be 27 states left light after
Wilson line breaking with the gauge quantum numbers under $G/D$ as if they
belonged to a complete $27$ of $E_{6.}$ This follows because for each
component of $R$ it is possible to pick a component of the discrete group
representation with the correct transformation property to satisfy eq(\ref
{bc}). Of course these residual light components will correspond to {\it %
different} multiplets on $M$ so their couplings will not be related by the
underlying gauge symmetry but the overall multiplet content does keep a
memory of the underlying symmetry.

There is another extremely important factor determining the low energy
spectrum. Wilson line breaking does not affect the index of the Dirac
operator \cite{gsw} which, in turn, does not depend on the representation of
the unbroken group $G/D$. As a result if there are $n_{l}$($n_{r}$) left-
(right-) handed chiral superfields transforming as $R$ on the manifold $M/D$ 
{\it without }Wilson lines there must be left light left-handed fields
making up $n=(n_{l}-n_{r})$ complete representations $R.$ As discussed
above, the components of $R$ having different transformation properties
under $U_{f}$ will correspond to different fields on $M$ having different
transformation properties under $D.$

The appearance of these $n$ chiral representations coming in complete $E_{6}$
representations, even in the theory in which the gauge group is broken,
gives a natural explanation for the pattern of states of the observed
families as we discussed in the previous section. In addition there will be $%
m_{ns}$ copies of $(27+\bar{27})$ corresponding to the discrete group
non-trivial representations. However note that there is one further
possibility for light fields not yet discussed. On $M$ we may also have
states which belong to the singlet representation of $D.$ From eq(\ref{bc})
we see that on $M/D$ the light states must transform trivially under the
Wilson lines and thus do not make up complete multiplets under $G$. These
are the states that give a natural explanation to the doublet-triplet
splitting problem because if the states left light are Higgs doublets we see
there are {\it no} associated colour triplets and the doublet-triplet
splitting problem never arises.

To illustrate this we consider specific examples of Wilson loops which are
relevant for realistic theories. We focus on Wilson loops embedded in an $%
E_{6}$ factor which contains an $SU(3)\otimes SU(2)\otimes U(1)$ subgroup.
This case has been explored in detail by Breit, Ovrut and Segre \cite{ovrut}%
. As long as the discrete group by which we mod out is Abelian ($Z_{N}$),
the Wilson loop can be written down as a linear combination of the elements
of the Cartan subalgebra $H^{i}$ of $E_{6}$ \cite{slansky} 
\begin{equation}
U=\exp (i\Sigma \bar{\lambda ^{i}}H^{i}).
\end{equation}
Using the method of Weyl weights, we determine the symmetry breaking
direction in the weight space which leaves the Standard Model gauge group
unbroken. We know that the elements of the mass matrix for gauge fields in
four dimensions are proportional to 
\begin{equation}
(\sum_{i}\bar{\lambda}_{i}c_{i})^{2},  \label{masses}
\end{equation}
where $\bar{\lambda}_{i}$ are dual components of the symmetry breaking axis,
and $c_{i}$ are Dynkin components of the root of a given generator (i.e. a
given gauge field).

We want to leave the Standard Model gauge group, $SU(3)\otimes SU(2)\otimes
U(1)$, unbroken so we need to have these gauge fields massless. The
corresponding $E_{6}$ roots are: (000001), (0100-10), (0-10011), for color
and $\frac{1}{2}$(10001-1) for $I_{3}^{W}$ axis. Using eq(\ref{masses})
shows that $\bar{\lambda}$ must have the following form: $[-c,c,a,b,c,0]$.
Because the Wilson line is a representation of the discrete group $Z_{N}$,
each of the numbers $a,b,c$ must be of the form 
\begin{equation}
k{\frac{2\pi }{N}},  \label{warunek}
\end{equation}
where $k=0,...,N-1$ (k can be a priori different for $a,b$ and $c$
respectively).

For $E_{6}$ the most convenient way of presenting the Wilson loop is to
write it in the maximal subgroup basis, $[SU(3)]^{3}$. This gives 
\begin{equation}
\left( 
\begin{array}{ccc}
1 &  &  \\ 
& 1 &  \\ 
&  & 1
\end{array}
\right) _{C}\times \exp i\left( 
\begin{array}{ccc}
-c &  &  \\ 
& -c &  \\ 
&  & 2c
\end{array}
\right) _{L}\times \exp i\left( 
\begin{array}{ccc}
a-b &  &  \\ 
& c-a &  \\ 
&  & b-c
\end{array}
\right) _{R}  \label{ug}
\end{equation}
Using the decomposition of the 27 dimensional $E_{6}$ fundamental
representation 
\begin{equation}
27=(1,\bar{3},3)+(3,3,1)+(\bar{3},1,\bar{3})=\chi +Q+Q^{c}
\end{equation}
we can determine how the Wilson line will act on different elements of the
spectrum. We have 
\begin{equation}
\chi \equiv \left[ 
\begin{array}{ccc}
H^{0} & H^{+} & e^{c} \\ 
\tilde{H}^{-} & \tilde{H}_{0} & \nu ^{c} \\ 
e^{-} & \nu & N
\end{array}
\right] \equiv \left[ 
\begin{array}{cc}
H_{u} & e^{c} \\ 
H_{d} & \nu ^{c} \\ 
\ell & N
\end{array}
\right] ;Q\equiv \left[ 
\begin{array}{c}
u \\ 
d \\ 
D
\end{array}
\right] \equiv \left[ 
\begin{array}{c}
q \\ 
D
\end{array}
\right] ;\;Q^{c}\equiv \left[ 
\begin{array}{c}
u^{c} \\ 
d^{c} \\ 
{D^{c}}
\end{array}
\right] \equiv \left[ 
\begin{array}{c}
q^{c} \\ 
{D^{c}}
\end{array}
\right]  \label{mcont}
\end{equation}
where we have shown the assignment of the first generation of quarks and
leptons and the fields $H_{u},\;H_{d}$ have the quantum numbers of the Higgs
supermultiplets in the MSSM and $D$ and $D^{c}$ are new colour  
triplets with the same charge as the $d$ quarks. The transformation
properties of these fields under the action of the Wilson line is given by
Table 1. 
\begin{table}[tbp]
\begin{center}
\begin{tabular}{||c|c|c|c|c|c|c|c|c|c|c||}
\hline
$\ell$ & $e^c$ & $\nu^c$ & $H_u$ & $H_d$ & $N$ & $q$ & $D$ & ${D^c}$ & $d^c$
& $u^c$ \\ \hline
$b$ & $a-b-2c$ & $-c-a$ & $a-b+c$ & $2c-a$ & $b-3c$ & $-c$ & $2c$ & $c-b$ & $%
a-c$ & $b-a$ \\ \hline
\end{tabular}
\end{center}
\caption{ Transformation of the components of a 27 dimensional $E_{6}$
representation under the Wilson line discrete group element.}
\end{table}
 From it we can readily determine for any model the conditions necessary to
leave a given particle massless.

\subsection{Light Higgs doublets.}

As we have discussed, after compactification there are $(n+m_{ns})$ complete
27s and $m_{ns}$ complete $\bar{27}$s and $m_{s}$ split multiplets. Let us
consider how this can lead to the light Higgs doublets needed in the MSSM.
Consider first the $(n+m_{ns})$ 27s. As may be seen from eq(\ref{mcont}) the
field content of a 27 dimensional representation of $E_{6}$ contains states
with the quantum numbers of the two Higgs doublets needed in the MSSM.
However it also contains the colour triplet states, $D$, the partners of the
Higgs multiplets making up the $5+\bar{5}$ $SU(5)$ vectorlike
representations. There must be a stage of spontaneous symmetry breaking
giving the $D$s a large $SU(3)\otimes SU(2)\otimes U(1)$ invariant mass.
This can arise through a coupling of $D,D^{c}$ to the $N$ field in eq(\ref
{mcont}), if $N$ acquires a large vev. In a GUT arranging for this to happen
while keeping their Higgs partners light is difficult because their
couplings are related by the GUT gauge symmetry. In the string case without
Grand Unification the situation is quite different because, as noted above,
the D components and the Higgs components do not come from the same 27 in
the theory before modding out. For this reason their couplings are not
related and it is quite possible for the $D$s to obtain a large mass
coupling to the N field while the Higgs remain massless because (presumably
due to a symmetry of the theory) they do not couple to the N field. The need
for such a symmetry is generic to supersymmetric extensions of the Standard
Model. In the MSSM the ``mu'' term in the superpotential $\mu H_{1}H_{2}$
must have the coefficient $\mu $ of order the electroweak breaking scale.
Since the unbroken Standard Model gauge group allows such a mass term its
smallness must be due to an additional discrete symmetry left unbroken by
the large scale breaking. We refer to this symmetry as the ``$\mu$'' symmetry.

A second possibility for generating a light Higgs arises because, as we have
already noted, the Wilson line breaking mechanism provides us a very natural
way of generating some extra states which come as incomplete representations
of the initial ($E_{6}$) gauge group. If there are any discrete group
singlet representations in the original manifold then necessarily there will
be such split multiplets. Indeed in Calabi Yau compactification we know
there is at least one pair of such fields transforming as $27+\bar{27}$, the 
$\bar{27}$ corresponding to the harmonic $(1,1)$ K\"{a}hler form so it is a 
{\it prediction} of the theory that there should be at least one multiplet
below compactification that does not fit into a complete $SU(5)$
representation! If the split multiplet should contain just the Higgs
doublets it solves the doublet triplet splitting problem. That this can
happen with Abelian Wilson loops follows from Tables 1 and 2. One may see
that the choice $b=a+c$, $a,c\ne 0$, $a\ne \pm c$, $a\ne 2c$ leaves only the
component $H_{u}$ light in the original 27 $(2,1)$ form and breaks the gauge
symmetry down to that of the Standard Model. Note that there is also left
light a Higgs doublet, $\bar{H}_{u}$, in the conjugate representation these
fields originating in the harmonic $(1,1)$ forms corresponding to the $\bar{%
27}$ representations on the original manifold \footnote{%
The case $b=3c,\;a=2c$ leaves both $H_a$ and $H_b$ light in the $(2,1)$ form
but has also a further pair of Higgs doublets in the conjugate
representation. Furthermore in this case the gauge group is $SU(3) \otimes
SU(2) \otimes SU(2) \otimes U(1) \otimes U(1)$. However if the $SU(3)
\otimes SU(2) \otimes U(1)$ singlet fields $\nu_R$ and $N$ acquire vevs the
gauge group will be further be broken down to just the Standard Model.
Indeed there exist compactifications in which these fields acquire Planck
scale vevs, effectively breaking the initial gauge symmetry to just $SU(5)$.}%
. The field $\bar{H}_{u}$ has the correct quantum numbers to play the role
of $H_{d}$ but, if so, there is an interesting implication for fermion
masses as discussed below. From eq(\ref{ug}) one may see that in this case
the group left unbroken by Wilson lines is $SU(3)\otimes SU(2)\otimes U(1)$
provided the discrete group is $Z_{N}$ with $N>3$. Thus we have a
straightforward mechanism capable of leaving just a doublet Higgs pair of
fields light after compactification. Note that any large scale breaking such
as is necessary to give mass to the vectorlike $5+\bar{5}$ representation
must {\it not} generate a mass term for $H_{u},\;\bar{H}_{u}$ otherwise the
mechanism for keeping Higgs doublets light fails. This should follow from
some symmetry of the theory. This is just the general ``$\mu$'' symmetry
discussed above for this case.

While the scheme just considered is very attractive it is {\it not} the only
natural way a light pair of Higgs doublets may occur. An alternative is for
the ($(2,1)$) split multiplet to contain just the $D$ states of eq(\ref
{mcont}) \footnote{%
This corresponds to the case $b=c$ in Table 2.}. We will denote these states
by $D_{S}$. The reason this apparently bizarre suggestion is viable follows
because these states, left light after compactification, may acquire a large
mass through a stage of spontaneous symmetry breaking by coupling to the $%
D^{c}$ components of one of the complete 27 dimensional set of fields needed
to accommodate the families as discussed above. Of course, just as in the
previous example, there will also be $\bar{D_{S}}$ states left light in the $%
(1,1)$ form and this should couple to the $D$ component of the 27. These
masses follow from the couplings 
\begin{equation}
<\phi >\bar{D}_{S}D+<N>D_{S}D^{c}
\end{equation}
where $\phi $ is the $E_{6}$ singlet field whose vev is responsible for
generating the masses in the additional vectorlike $m_{ns}(27+\bar{27})$
fields via the coupling $\phi 27\bar{27}$. In this case the ``$\mu$''
symmetries of the theory should {\it forbid} the $N5\bar{5}$ mass term of
the non-MSSM states in one of the family 27s. The net result is that, in
addition to the Standard Model states, only the Higgs doublets $H_{u}$ and $%
H_{d}$ are left light from this family 27. We shall discuss in the next
section the different phenomenological implications of the three mechanisms
we have identified capable of giving rise to the light Higgs doublets.

\section{Unification predictions}

As we have discussed, even in string theories without a Grand Unified group
below the compactification scale, the effect of the underlying extended
gauge symmetry remains in the light spectrum. Even if the gauge group is
just that of the Standard Model the gauge couplings are related and the
unification scale determined close to the compactification scale. Thus such
theories can lead to predictions for the low-energy gauge couplings of a
similar nature to those obtained in Grand Unification. Particularly since
such predictions provide the main evidence for a stage of Grand Unification,
it is of interest to examine these predictions in some detail, for the
string unification case in which the gauge group is not Grand Unified below
the compactification scale.

Thus we consider level-1 string theories in which the couplings are related
at the unification scale in the same way as they would be in an $SU(5)$
theory but, due to Wilson line breaking, the gauge group below the
compactification scale is just that of the Standard Model. The multiplet
structure is just that of the standard model plus the additional heavy
states which acquire $SU(3)\otimes SU(2) \otimes U(1)$ invariant masses {\it %
below} the unification (or compactification) scale. Thus we must allow for
additional multiplets with Standard Model gauge content equivalent to
``vectorlike'' representations coming in $(5+\bar{5})$ or $(10+\bar{10})$
representations of $SU(5)$ plus any states associated with the appearance of
the Higgs ``split'' multiplets in the MSSM. The latter fall into two main
categories. In the more ``conventional'' case there are just two split
multiplets containing the MSSM Higgs supermultiplets and so there are no
further massive multiplets to consider. In the unconventional case the pair
of light Higgs at low energy comes from a $5+\bar{5}$ plus a $D_S+D^c_S$
split multiplet, the new states obtaining intermediate scale masses. The
third possibility is that the Higgs came from a $5$ associated with one of
the families and in this case there will be additional $D+D^c$ states at the
intermediate scale. The implications for unification of couplings of this
case is similar to the unconventional case as discussed below but, as we
shall see, the implications for fermion masses do vary.

Let us first consider the ``conventional'' split multiplet case. In addition
to the MSSM states, the only flexibility in the spectrum is the appearance
of additional vectorlike states filling out complete 5 or 10 dimensional $%
SU(5)$ representations together with some Standard Model singlet states. 
The expectation is that these vectorlike states will acquire large masses. Since they come in vectorlike combinations, we can write down $SU(3)\otimes SU(2)\otimes U(1)$ invariant mass terms $\mu _{\psi }\psi \bar{\psi}$ where the mass $\mu _{\psi }$ can 

be far above the electroweak breaking scale. This term will arise from a stage of spontaneous symmetry breaking through the coupling $\lambda _{\psi }\phi \bar{\psi} \psi$, when the $SU(3)\otimes SU(2) \otimes U(1)$ invariant scalar field $\phi$ acquires 
a vacuum expectation value (vev). There is a very natural explanation for the origin of this vev, because the mass squared of the fields $\phi$ will be driven negative by the radiative corrections involving the coupling $\lambda _{
\psi }$, in the usual radiative breaking 
mechanism \cite{marcodum}.

The
unification predictions of this class of model differs from the minimal
unification predictions due to the radiative effects of these new states
which, while expected to be heavy due to intermediate scale breaking are
lighter than the unification scale. The phenomenological effects of such a
non-standard spectrum has been discussed in detail in \cite{marcodum} 
This analysis uses the measured values of $\alpha_{1,2}(M_Z)$ to predict the value of $\alpha_3(M_Z)$ in terms of the unified coupling, {\it assuming} string threshold corrections are negligible. At one
loop the predictions for the unification scale and $\alpha _{3}(M_{Z})$ do
not change. This follows because complete SU(5) multiplets give the same
additive contribution to the $SU(3)$, $SU(2)$ and $U(1)$ beta functions. The
predictions for the unification scale and $\alpha _{3}(M_{Z})$ follow from
the differences of the beta functions, hence this result. However the
magnitude of the gauge coupling at unification does change due to the extra
light matter, causing it to increase. At two loop order there are two
effects. The first is the usual two-loop contribution to the beta functions.
The second is the radiative splitting of the masses of the new intermediate
scale mass states. This splitting occurs at one loop order but its effect on
the gauge couplings occurs only via the one loop beta function and hence the
net effect is at two loop order. Remarkably the two effects largely cancel 
\cite{marcodum,shifman} leaving the predictions for $\alpha _{3}(M_{Z})$ and $% 
M_{X}$ very close to those of the MSSM even in the case there is a large
number of additional vectorlike states left light after
compactification. At
this order the predictions depend only on the combination $p=n_{5}+3n_{10}$
where $n_{5}$ and $n_{10}$ denote the number of additional 5s and $n_{10}$
the number of additional 10s. For $\alpha_i (M_{X})$ in the perturbative
domain the value of $\alpha _{3}(M_{Z})$ and $M_{X}$ are both increased
somewhat from the MSSM. For example for $p=30$ and $\alpha_i (M_{X})=1$, just
at the limit of perturbation theory, $\alpha _{3}(M_{Z})=0.13$ and $M_{X}$
increases by a factor 3.5. For $p=5$, $\alpha _{3}(M_{Z})=0.132$ and $M_{X}$
increases by a factor 7.5. The value of $\alpha _{3}$ is uncomfortably large
compared to the experimental mean of $0.118\pm 0.003$, but the increase in $%
M_{X}$ for $p=5$ takes it closer to the (weakly coupled) heterotic string
value.

We turn now to the ``unconventional" split multiplet case in which, below
the compactification scale, there are additional light colour triplets. At
one loop order this affects the running of the couplings and hence the
unification predictions 
\begin{equation}
\alpha_{i}^{-1}(Q)=\alpha_{GUT}^{-1}+{\frac{b_i }{2 \pi}} \log {\frac{Q }{M}}
+ {\frac{b_{i}^{^{\prime}} }{2 \pi}} \log {\frac{M }{M_{X}^{\prime}}}
\label{coup}
\end{equation}
where $M_X^{\prime}$ is the unification scale and $M$ is the mass of the
additional colour triplet states. Below $M$ the one loop beta function, $b_i$%
, is just that of the MSSM. Above $M$ the beta function changes to $b_{i}^{\prime}$. 

We wish to determine the change in the unification scale and strong coupling in this scheme relative to the MSSM in which
\begin{equation}
\alpha_{MSSM,i}^{-1}(Q)=\alpha_{MSSM,GUT}^{-1}+{\frac{b_i }{2 \pi}} \log {\frac{Q }{M_X}} 
\label{coup1}
\end{equation}
where $M_X$ is the unification scale in the MSSM. In both cases $\alpha_{1,2}(M_Z)$ are input as their measured values. The unification scale is found from the relative evolution of $\alpha_{1,2}$. In the model with $D$s  above M the relative evolution of
 $\alpha_2$ speeds up 
causing $\alpha_{1,2}$ to meet sooner {\it lowering} the unification scale. On the other hand due to the $D$s, the relative evolution of $\alpha_3$ and  $\alpha_2$ 
is decreased 
above M so the value of $\alpha_3(M_Z)$ must be {\it reduced} to allow the couplings to meet. Quantitatively, combining eqs(\ref{coup},\ref{coup1}), we find 
\begin{equation} 
\log \left( {\frac{M_Z }{M_X}} \right)=\log \left( {\frac{M_Z }{%
M}} \right) + {\frac{b_{2}'-b_{1}' }{b_{2}-b_{1} }}
\log \left( {\frac{ M} {M_{X}'}} \right),
\end{equation}
or 
\be
\label{newsc}
\log \left( {\frac{M}{M_X}} \right)={\frac{%
b_{2}'-b_{1}'}{b_{2}-b_{1} }} \log \left( {\frac{M} 
{M_{X}'}} \right).
\ee
and
\be 
\label{nowed}
\Delta \alpha _{3}^{-1}=-{\frac{1}{b_{1}'-b_{2}'}}\left[
(b_{2}'-b_{3}') \Delta b_{1} + (b_{3}'-b_{1}') \Delta b_{2}
+(b_{1}'-b_{2}') \Delta b_{3} \right] 
{\frac{1}{2\pi }}\log \left( {\frac{M}{M_{X}}}\right),
\ee
where $\Delta b_i$ are definied to be
\be 
\Delta b_i = b_{i}' - b_i.
\ee 

The MSSM prediction for $\alpha_3$ is approximately 0.006 larger than the
experimental measurement. This discrepancy can be eliminated by choosing $M_X'$ appropriately. This gives
\begin{equation}
\frac{M_X'}{M_X}\approx 0.9 
\end{equation}
Note that the effect of this scheme is largely confined to a change in $\alpha_3$, the change in the unification scale being very small.

So far we have neglected the possible effect of additional vectorlike states
on our prediction. At one loop these do not change the expectation for the
unification scale or the strong coupling. However they do cause the coupling
at unification to increase, increasing the potential importance of two loop
effects. These are straightforward to compute and have been considered
in detail in \cite{marcodum} (the effect is that
both $M_{X}$ and $\alpha _{3}$ increase). In this case however the spectrum
including the $D$ quark split multiplets reduces $\alpha _{3}$ at one loop
so the effects together lead to a net reduction which may be adjusted to give the experimentally measured value through choice of $M_X$.
However, as we have seen, the new one loop effects leave $M_{X}$ essentially
unchanged so, for it,  the net effect is an increase in the value of $M_{X}$ from the
MSSM value. For a perturbative value of the unified coupling the increase in 
$M_{X}$ is only a factor of (3.5-7.5), the upper value reducing, but not
eliminating, the discrepancy with the string prediction in eq(\ref{su}).  

We have considered the gauge unification predictions for two of the three
possible mechanisms we identified as being able to give light Higgs
doublets. The third possibility is that the Higgs came from a $5$ associated
with one of the families and in this case there will be an additional $D+D^c$
states at the intermediate scale. This possibility is intermediate between
the first two because in the conventional split multiplet case there are no
additional states and in the unconventional split multiplet case there are
two pairs of $D+D^c$ states at the intermediate scale. Thus the prediction
for this third possibility follows that of the unconventional case provided
one adjusts the intermediate mass scale to take account of the reduced
number of $D$ states. In this latter case there is a further ambiguity in
the intermediate scale spectrum corresponding to the split multiplets which
must acquire intermediate scale masses. We will not consider their effect
here.

\subsection{Fermion masses}

We close this section with a brief discussion of the expectation for fermion
masses in the class of string theory of interest here. As we noted above the
case of unification at strong coupling is particularly interesting and is
motivated by the appearance of the additional vectorlike states in the
string theory. In this case the ratios of fermion masses are determined by
infra-red fixed points. The details depend on which Yukawa couplings are
allowed by the symmetries of the model and have been studied in \cite
{marcodum}. There it is found, for example, that the ratio $m_b/m_{\tau}$,
may be driven to an initial value close to 1 leading to an excellent
prediction for the $b$ mass at low energies. This prediction does {\it not}
rely on a GUT relation between the couplings. Rather it follows because the
multiplet structure carries the underlying GUT structure allowing for a
similar set of couplings involving the $b$ and the $\tau$ and thus leading
to fixed points for these couplings which are nearly the same for the $b$
and the $\tau$.

However, there may be important differences in the prediction for the quark
masses in the various versions of our string compactifications. This follows
from the structure of the light Higgs sector discussed above. Let us
consider both the conventional and unconventional split multiplet cases. In
this the Higgs doublets $H_{u}$ and $\bar{H}_{u}$ are $(2,1)$ and $(1,1)$
harmonic forms respectively\footnote{%
We assume the chiral multiplet excess is in the $(2,1)$ sector. If it is in
the $(1,1)$ sector the same discussion applies with $(2,1)$ and $(1,1)$
interchanged.}. However the chiral matter fields are $(2,1)$ harmonic forms.
The couplings leading to fermion masses come from the coupling of three $%
(2,1)$ forms there being no coupling of two $(2,1)$ forms to a $(1,1)$ form.
Thus we have the prediction that only the up quark masses are non-zero.
While this is a good starting point, explaining why the top mass is much
larger than the bottom (and $\tau $) mass, it is only the first
approximation. The down quark and lepton masses will be generated through
mixing of the ($(1,1)$) $\bar{H}_{u}$ with the ($(2,1)$) $H_{d}$ states in
the spectrum below the compactification scale. As discussed above these
states appear as part of the fields making up the 27 dimensional
representations the families belong to and as part of the additional
vectorlike states filling out complete $27$ plus $\bar{27}$ representations.
This mixing is generated through the couplings needed to give the non-MSSM
states a mass. Suppressing Yukawa couplings we have 
\begin{equation}
\phi H_{u}\bar{H}_{u}+NH_{u}H_{d}
\end{equation}
If $\phi $ and $N$ acquire vevs the combination $<\phi >\bar{H}_{u}+<N>H_{d}$
is heavy leaving the combination $<N>\bar{H}_{u}-<\phi >H_{d}$ light. The
component $H_{d}$ {\it can} couple to down quarks (and leptons) allowing for
the generation of down (and lepton) quark masses. The important point is
that the ratio of up to down quark masses is determined by the ratio $<\phi
>/<N>$ offering a simple explanation for the difference between the top and
bottom (and $\tau $) masses.

Finally we consider the third case in which the Higgs originate in one of
the family 27s, {\it i.e.} both Higgs are (2,1) forms. In this case both can
couple directly to the quarks and so there is no immediate reason for the
top Yukawa to be enhanced relative to the bottom Yukawa. This case fits most
easily with the large $\tan \beta $ solution in which the difference between
the top and bottom masses follows because of the asymmetry between the two
Higgs vacuum expectation values.

\section{Summary and Conclusions}

While there has been much progress in understanding and constructing four
dimensional string theories there has been relatively little progress in
classifying their phenomenological implications. This is mainly because the
vast number of candidate string vacua makes it difficult to draw general
conclusions. In this paper we have attempted to make progress in this
direction by identifying what we consider to be a particularly promising
class of string compactifications and studying some general phenomenological
features they display. The class of string theories we consider is based on
compactification of the level-1 heterotic string with Wilson loops. This
gives a low-energy structure remarkably close to that observed. The families
fill out complete $SU(5)$ representations even though the Gauge Group is not
Grand Unified. The quarks and leptons must belong to the fundamental
representation of the Standard Model group, higher representations are not
allowed. The gauge couplings have the $SU(5)$ relations at the string
unification scale giving rise to the good unification predictions for the
couplings at low energies even though there is no stage of Grand
Unification. The problem encountered in GUTs of splitting the light Higgs
doublets from their colour triplet partners is evaded because the Higgs
states are not related by a GUT to colour triplets.

Given this motivation we considered in some detail the phenomenological
implications of this class of theory. The first important observation is
that the spectrum will be that of the MSSM plus additional states which are
left light at the compactification scale but which will acquire mass at an
intermediate scale. The nature of this additional matter is quite
constrained. The light multiplet structure after compactification consists
of light left-handed chiral superfields which have the same multiplet
content as $(n+m_{ns})$ complete $27$s and $m_{ns}$ complete $\bar{27}$s of $%
E_{6}$ even though the gauge group is just that of the Standard Model. In
addition there are $m_{s}$ split multiplets consisting of those components
of the $(27+\bar{27})$ which are singlets under the Wilson line group
elements.

We identified three mechanisms for giving rise to the light Higgs doublets
of the MSSM and analyzed their predictions for the unification scale and the
strong coupling, {\em assuming} string threshold corrections are
negligible. The first has  the light Higgs doublets coming from the
split multiplets. It gives a prediction for $\alpha _{3}(M_{Z})$ larger than
that of the MSSM, perhaps unacceptably so for the case when unification
occurs for a large value of the gauge coupling. In this case $M_{X}$ is also
increased, by a factor (3.5-7). The second mechanism to get light Higgs
doublets also involved the split multiplets but in this case they contain $D$
quarks. Thus the spectrum above the intermediate scale is modified. This has
the interesting effect of lowering $\alpha _{3}$, easily accommodating the
experimental value. For the case of large unified coupling $M_{X}$ is still
increased by a factor (3-6). The third case in which the light Higgs
originate from one of the family multiplets gives predictions quite similar
to this case.

We find it remarkable how close to the Standard Model structure this class
of string theory is. Moreover it illustrates quite clearly how the
unification predictions which have led many to conclude there is a stage of
Grand Unification may follow equally well without Grand Unification and
without many of the complications that are required to build a viable Grand
Unified theory. The precise predictions are sensitive to the mechanism
giving light Higgs doublets and it is interesting that two of the possible
mechanisms offer an explanation for the discrepancy between the MSSM\
prediction for the strong coupling and its precision measurement. String
unification without Grand Unification leads to a quantitative test of the
idea of unification including gravity for the gauge unification scale of the
Standard Model couplings is also the string unification scale. The initial
comparison is encouraging. In the case of unification at large coupling the
gauge unification scale is found to be $(0.9-1.8)10^{17}{\rm GeV}$ to be
compared with the string prediction following from eq(\ref{su}) of $%
1.5\,10^{17}{\rm GeV}$ in the 10D weakly coupled heterotic string case and
somewhat less in the strongly coupled case.

W.P. is grateful to B.C. Allanach, H. Dreiner and S. Pokorski for help and
would also like to thank V.M. Annett, D. Ghilencea and K. Smaga for many
stimulating discussions. W.P. gratefully acknowledges the financial support
of the Oxford Overseas Bursary.

\end{document}